\documentclass[openacc]{rstransa}%%%%where rstrans is the template name
\usepackage[numbers, compress]{natbib} % round optional

\usepackage{booktabs,tabularx,array}
\newcolumntype{Y}{>{\raggedright\arraybackslash}X}   % your left-ragged X
\newcolumntype{C}{>{\centering\arraybackslash}X}     % centered X

\titlehead{Review}

\begin{document}

%%%% Article title to be placed here
\title{Linking Solar Magnetism,  Extreme Solar Particle Events and Stellar Superflares}

\author{%%%% Author details
Valeriy Vasilyev$^{1}$, Natalie Krivova$^{1}$ and Ilya Usoskin$^{2,3}$}

%%%%%%%%% Insert author address here
\address{$^{1}$Max Planck Institute for Solar System Research, Justus-von-Liebig-Weg 3, 37077 G\"ottingen, Germany\\
$^{2}$Space Physics and Astronomy Research Unit and Sodankyl\"a Geophysical Observatory, University of Oulu, 90014 Oulu, Finland\\
$^3$Institute for Space-Earth Environmental Research, Nagoya University, Furo-cho, Chikusa-ku, Nagoya 464-8601, Japan}

%%%% Subject entries to be placed here %%%%
\subject{xxxxx, xxxxx, xxxx}

%%%% Keyword entries to be placed here %%%%

%%%% Insert corresponding author and its email address}
\corres{Valeriy Vasilyev\\
\email{vasilyev@mps.mpg.de}}

%%%% Abstract text to be placed here %%%%%%%%%%%%
\begin{abstract}
The magnetic field of the Sun drives a wide range of eruptive phenomena, from small-scale nanoflares to large flares and coronal mass ejections (CMEs).  While direct observations of solar activity cover only the past few decades, indirect evidence indicates that the Sun can occasionally produce events orders of magnitude stronger than any recorded ones in the modern era. Two complementary lines of evidence exist. First, extreme solar particle events (ESPEs) have been inferred from prominent spikes in cosmogenic isotope concentrations preserved in precisely dated natural archives such as tree rings and ice cores over the past 15 millennia. Second, high-precision space-borne photometry has revealed superflares on thousands of stars similar to the Sun. Whether these solar and stellar extremes are physically related remains an open question. We summarise the present state of understanding and discuss physical mechanisms that may link them. Although superflares and ESPEs are both extremely energetic manifestations of magnetic energy storage and release, their relationship does not appear to be one-to-one. Their occurrence and energetics likely depend on how magnetic flux and topology govern the partitioning of released energy between radiation, mass ejection, and particle acceleration.
\end{abstract}
%%%%%%%%%%%%%%%%%%%%%%%%%%%

\begin{fmtext}
\end{fmtext}
\maketitle
\section{Introduction}
\label{S:Intro}
The Sun has always been of great interest to humanity, as the ultimate source of light, warmth, and life on Earth.
Its daily and seasonal regularity provided the basis for agriculture, timekeeping, navigation, and eventually the development of human culture.
The apparent constancy of the Sun led to the long-standing perception of the Sun as a perfect, unchanging body.
This perception began to change when Galileo Galilei and Christoph Scheiner independently observed sunspots~-- dark blemishes that challenged the doctrine of solar perfection. 
The discovery of the approximately 11-year sunspot cycle by Schwabe in 1843, followed by large solar storms such as the so-called Carrington event of 1859, finally revealed that the Sun is not static but an active, magnetically variable star capable of dramatic outbursts.
The Carrington event at the beginning of September 1859 was the first observation of a solar flare and a related geomagnetic superstorm \cite{tsurutani2003}.
Later, it was discovered that solar flares can be accompanied by bulk acceleration of energetic particles \cite{forbush46}, currently known as solar particle events (SPEs) or, if they can be detected by ground-based detectors, ground-level enhancements (GLEs) \cite{poluianov17}. 

The Sun, its variability and activity have been extensively studied over the past decades by accumulating a large observational basis and advances in theories and models of sporadic eruptive events \cite{benz17,webb12}.
Yet, despite this progress, two major discoveries in 2012 revealed that our
understanding of solar activity is still incomplete, and that the Sun has the potential to produce more extreme eruptive events than previously recognised.

The first was the detection of a pronounced spike in radiocarbon $^{14}$C measured in tree rings, dated to the year AD 775 \cite{miyake12}.
This spike was shown to have been caused by solar energetic particles \cite{usoskin_775_13}.
Further similar events were later identified over the past 15 millennia \cite{Cliver2022,usoskin_SSR_23}, which implies the discovery of a new %previously unknown 
type of solar events: extreme solar particle events (ESPEs)~-- see Sect.~\ref{S:ESPE}. 
The second discovery was the detection of superflares on stars similar to the Sun in high-precision photometric surveys \cite{Maehara2012}, see Sect.~\ref{S:Stellar}.
Although superflares were known before, they were detected on magnetically very active stars, and their discovery on Sun-like stars changed the landscape significantly, suggesting that even our Sun can potentially unleash such strong eruptions. 
Although the possibility that superflares can occur on the Sun is broadly acknowledged \cite{Cliver2022}, the estimates of the strength and occurrence rates of superflares when projected to the Sun remain somewhat controversial~-- see Sect.~\ref{S:Stellar}.
Both these types of events are related to extremely strong eruptive releases of magnetic energy in the solar/stellar atmospheres.
%surface and corona, 
However, the relationship between them is unclear.
It is still unknown whether they are two sides of the same coin or whether they are independent of each other. 
The present-day knowledge about the relation between solar flares and SPEs via solar surface magnetism is summarised in Sect.~\ref{S:Solar}.

Here, we present an overview of the current state of the art in understanding extreme solar/stellar eruptive events and their manifestations. 

%=======================
\section{Extreme Solar Particle Events in cosmogenic isotope records}
\label{S:ESPE}

\subsection{Discovery and interpretation}
\label{sec:ESPE-discover}

The fact that the Sun can accelerate energetic charged particles, sporadically producing their fluxes detectable on the ground, was discovered in 1942 \cite{forbush46}.
Since then, thousands of sporadic SEP events have been detected, mostly in space, as characterised by a soft energy spectrum dominated by lower-energy ($<$100 MeV) protons \cite{desai_LR_16}. 
Sometimes, SEPs may have a harder spectrum, up to several GeV of protons' energy, and a high flux to be registered on the ground above the background of omnipresent galactic cosmic rays (GCRs), forming a subclass of SEP events, called GLE \cite{poluianov17}.
At present, 76 GLEs are known over about 80 years of instrumental observations, with the related information collected in the International GLE Database \footnote{\url{https://gle.oulu.fi}}.
The strongest recorded GLE took place on 23 February 1956 (GLE\#5) with the peak intensity reaching 4500\% of the GCR background for several hours \cite{usoskin_1956_20}.
The second-highest GLE was GLE\#69 on 20 January 2005 with an enhancement of several thousand per cent in polar neutron monitor count rate.
These GLEs were considered the worst-case SEP events at the edge of the Sun's ability to accelerate SEPs.

\begin{table}[t]
 \caption{Cosmogenic isotopes used in studies of ESPEs. The columns represent the isotope, its half-life, main production reaction, the main target nuclei in the atmosphere, and the effective energy of their production by SEPs \cite{koldobskiy22}, and natural archive.}
    \centering
    \begin{tabular}{cc|ccc|c}
    \hline
    Isotope & Half-life & Reaction & Target & $E_{\rm eff}$ (MeV) & Archive\\
    \hline
    $^{14}$C radiocarbon & 5730 yr & `neutron capture' & N &  $234\pm 18$ MeV & tree rings\\
    $^{10}$Be & 1.39 Myr & spallation & N, O & $236\pm 16$ MeV & ice cores\\
    $^{36}$Cl & 300 kyr & spallation & Ar & $60\pm 7$ MeV & ice cores\\
    \hline
    \end{tabular}
    \label{tab:IU-isotopes}
\end{table}

In 2012, an increase of $\approx$12\textperthousand\, in $\Delta^{14}$C has been discovered in a tree ring of a Japanese Cedar dated to year AD 775 \cite{miyake12}. 
Due to the attenuation effect of the carbon cycle (e.g., \cite{bard97}), this seemingly small enhancement implies an enormous, nearly four-fold, very short production spike of $^{14}$C above the cumulative annual GCR background.
Now, such strong and short spikes in radiocarbon production are collectively known as `Miyake events'. 
The initial interpretation of this increase involved exotic objects such as a nearby supernova explosion \cite{miyake12}, or a Galactic gamma-ray burst \cite{pavlov13}.
However, it was soon proven that the only feasible source of the Miyake event is related to an extreme solar energetic particle event (ESPE) produced by the Sun \cite{usoskin_775_13}.

Cosmogenic isotopes are produced by energetic particles (cosmic rays) in the Earth's atmosphere in a sequence of nuclear collisions and consequently stored in natural, independently dateable archives such as tree rings or ice cores \cite{beer12}.
Accordingly, the amount of cosmogenic isotopes recorded in the archive for some period reflects the flux of cosmic rays impinging on Earth and thus, solar activity in the past \cite{usoskin_LR_23}.
The most useful cosmogenic isotopes are $^{14}$C or radiocarbon, $^{10}$Be and $^{36}$Cl -- see table~\ref{tab:IU-isotopes}.
The isotopes are continuously produced by a slightly variable flux of GCR, but sometimes, extremely high fluxes of SEPs can produce clearly detectable spikes in the isotope production or Miyake events \cite{miyake19}. 
The detection of a spike in one natural archival record makes it a candidate for an ESPE. 
To be qualified as an ESPE, the event needs to be independently confirmed by several groups analysing different archival records.

At present, there are seven ESPEs and two candidates known, covering the period of 15 millennia back \cite{usoskin_SSR_23,golubenko25}.
This makes the mean occurrence rate of these events roughly once per 1500 years, but the occurrence is irregular and seemingly sporadic.
The strongest known Miyake event took place in 12350 BC with the $\Delta^{14}$C increase of 38\textperthousand \cite{bard23}, with the corresponding ESPE being 18\% stronger than that of the first discovered event (AD 774).

\begin{figure}[t]
    \centering
    \includegraphics[width=1.0\linewidth]{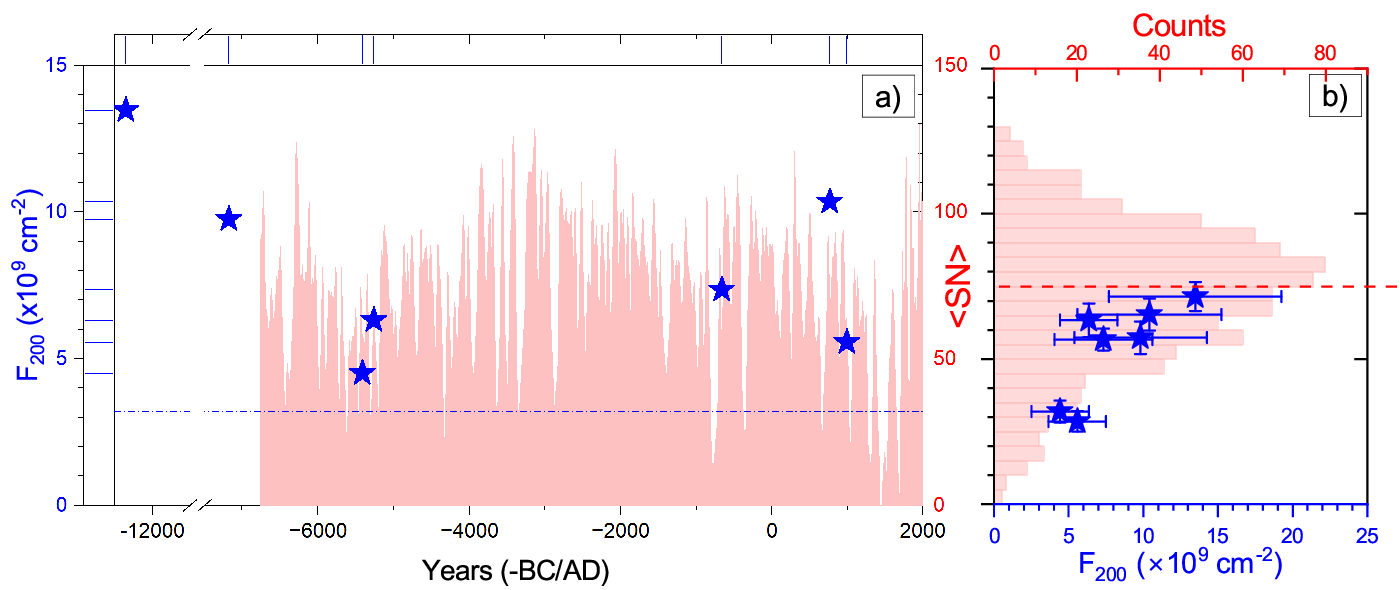}
\caption{\textbf{ESPE and sunspot activity.} 
\textbf{(a)} Time profile of decadal sunspot numbers $\langle{\rm SN}\rangle$ (red-filled curve, right axis) reconstructed from cosmogenic isotopes \cite{wu18}; Occurrence time of known ESPEs (blue stars, left axis, blue ticks on the left and top) quantified in the fluence (event-integrated flux) of SEPs with energy above 200 MeV, $F_{200}$ (data from \cite{golubenko25} and \cite{koldobskiy25}). Error bars are shown in panel (b).
The blue dash-dotted line denotes an estimate of the mean annual fluence $F_{200}=3.2\cdot 10^6$ cm$^{-2}$ over the past solar cycle 24 (2009\,--\,2019) \cite{raukunen22}, multiplied by 1000.
Note the break in the time axis.
\textbf{(b)} Distribution histogram of the $\langle{\rm SN}\rangle$ values shown in panel (a) (red horizontal bars, top and left axes); ESPE strength ($F_{200}$, bottom axis, blue stars with 1$\sigma$ error bars) as a function of $\langle{\rm SN}\rangle$ (left axis) during the decade of the ESPE occurrence.
The horizontal red dashed line denotes the median of the $\langle{\rm SN}\rangle$ distribution.
A similar figure but for $F_{\rm 800}$ can be found in \cite{koldobskiy25}.}
    \label{fig:IU-ESPE}
\end{figure}

\subsection{Reconstruction of parameters}

From a spike in the measured cosmogenic isotope concentration (Miyake event), the total number of isotope atoms produced in the Earth's atmosphere by an ESPE can be estimated, without any direct information on the energy of SEPs.
Such estimates are shown in Figure~\ref{fig:IU-ESPE} where the events' strengths are quantified in the integral fluence (event-integrated flux) of SEPs with energy above 200 MeV, $F_{200}$.
A strong ESPE can irradiate Earth with the fluence of SEPs exceeding the cumulative SEP fluence over a millennium of normal activity as observed during the previous cycle 24 (denoted by the blue dash-dotted line in Figure~\ref{fig:IU-ESPE}a).

\begin{figure}[t]
    \centering
    \includegraphics[width=0.8\linewidth]{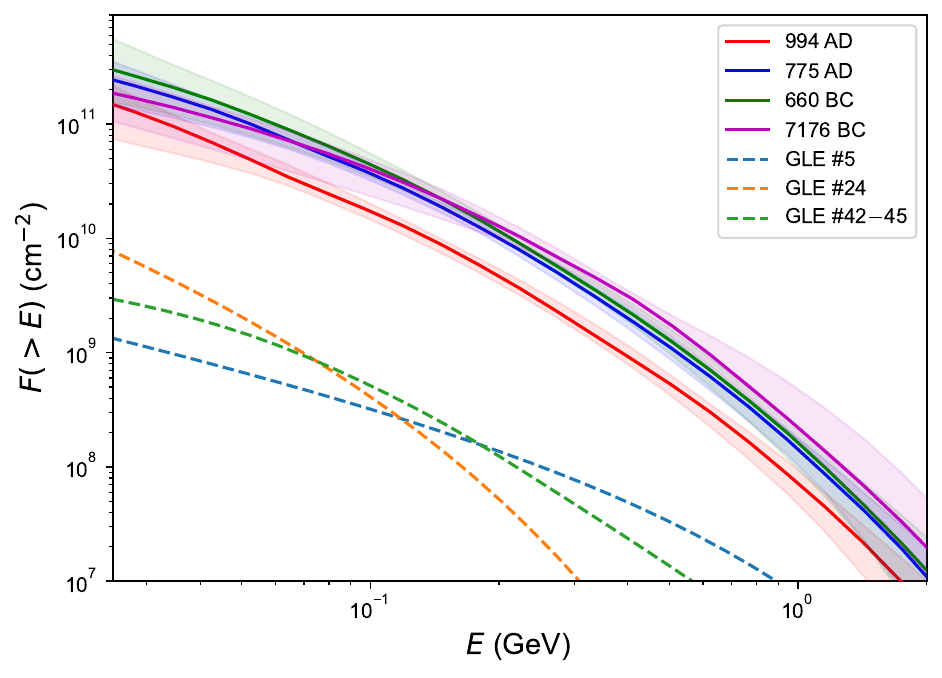}
\caption{\textbf{Integral energy spectra of ESPEs and strongest directly observed SEP events.} 
Solid curves with shaded areas denote multi-proxy reconstructions, with 68\% confidence intervals, of the reconstructed ESPE fluences. 
Dashed curves denote the integral spectra for three very strong SEP events: the hard-spectrum GLE\#5 (23-Feb-1956), a soft-spectrum
GLE\#24 (04-Aug-1972) and a typical-spectrum series of GLE\#42–45 (October–November 1989), as denoted in the legend.
Figure adopted from \cite{usoskin_LR_23}.}
    \label{fig:IU-spec}
\end{figure}

Different cosmogenic isotopes have different production cross-sections and, thus, different effective energies of their production by energetic particles in the atmosphere (Table~\ref{tab:IU-isotopes}).
Accordingly, the energy spectra of the primary particles can be parametrically estimated using a multi-proxy method, where the values of the parameters of a prescribed spectrum shape of ESPE are found that best fit the measured increases in different cosmogenic records for the same event \cite{mekhaldi15,ohare19,koldobskiy23}.
Of special importance is the isotope of $^{36}$Cl, which is more sensitive to low-energy particles, and the ratio of $^{36}$Cl to $^{10}$Be, measured preferably in the same ice core, provides an estimate for the softness/hardness of the energy spectrum \cite{mekhaldi15}. 
The ESPE spectra were found to be similar to those of recent strong events, but two orders of magnitude higher (Figure~\ref{fig:IU-spec}).
This multi-proxy method unambiguously confirmed that the observed Miyake events were caused by ESPEs, excluding such sources as gamma-ray bursts or GCR \cite{usoskin_775_13}.

A parametric multi-proxy reconstruction of the energy integral spectra for several strong ESPEs is shown in Figure~\ref{fig:IU-spec}.
The reconstruction was based on a simultaneous fitting of the measured records for the three isotopes (Table~\ref{tab:IU-isotopes}) by scaled spectra of a variety of observed SEPs (see details in \cite{koldobskiy23}).
The scaling factor was the only free parameter of the fit.
As seen, the reconstructed spectra (solid lines) generally agree with typical SEP spectra (dashed lines) but are roughly two orders of magnitude higher.
The three strong ESPEs (775 AD, 660 BC and 7176 BC) appear fully consistent with each other, implying that their origins were similar.
The spectrum of a smaller ESPE of 994 AD had a similar shape but was slightly lower.
This suggests that ESPEs are the most likely (and the only feasible in the framework of the present knowledge) source of the known Miyake events with the energy spectra typical for the modern SEP events, but several orders of magnitude more intense.

There is another, very indirect way to evaluate the average flux of ESPEs on a very long timescale.
It is based on measurements of cosmogenic isotopes in lunar soil and rocks.
Since the Moon does not have its own magnetic field or atmosphere, energetic particles can freely impinge on its surface, producing isotopes inside its soil (regolith) or rocks, with the mean depth of the production defined by the particles' energy.
Since the lower-energy range of energetic particles ($<$80 MeV) is dominated by SEPs over GCRs, isotopes in the upper, few-cm-thick layer of lunar soil/rock are produced predominantly by SEPs and remain there until their decay. 
Several samples of lunar rocks and regoliths were brought to Earth by \textit{Apollo} missions and measured in laboratories (e.g., \cite{nishiizumi84,nishiizumi09}). 
By measuring the concentration of cosmogenic isotopes as a function of depth in lunar rocks, one can estimate an average spectrum of SEPs on the lifetime of the isotope \cite{poluianov18}.
This was made, using $^{26}$Al (half-life 717 kyr) measured in lunar rocks.
Recent estimates \cite{poluianov18} suggest that the average flux of SEP with energy $>$30 MeV, $F_{30}$, was $38\pm 7$ cm$^{-2}$ s$^{-1}$, while the $F_{30}$ flux directly measured over the solar cycle 24 (2009\,--\,2019) was lower, about 8 cm$^{-2}$ s$^{-1}$ \cite{raukunen22}.
Meanwhile, the long-time averaged $F_{30}$ flux corresponding to ESPEs is estimated as 10\,--\,40 cm$^{-2}$ s$^{-1}$ \cite{usoskin_AAL_23}, making all numbers consistent. 
It was concluded, based on the lunar data, that ESPEs contribute 40\,--\,80 \% to the very long-term SEP flux.
This suggests that ESPEs are rare but regular phenomena on the Sun.

\subsection{Summary}

As we know from indirect proxies (cosmogenic isotopes), the Sun can rarely, roughly once per one--two millennia, produce extreme events, ESPEs. 
The fluence of lower-energy particles during such ESPEs is enormous and can exceed the cumulative SEP fluence accumulated over a millennium.
This is independently confirmed by isotopic measurements of lunar regoliths and rocks that ESPEs can contribute roughly half of the very-long-term flux of SEPs.
Acceleration of so high fluxes of energetic particles by the Sun goes beyond the existing knowledge of solar eruptions and requires the development of new models capable of simulating such processes \cite{Cliver2022,usoskin_SSR_23}.

At present, nine extreme events and candidates are known over 15 millennia backwards.
The strongest ESPEs are those of 12350 BC and 774 AD, and this likely indicates the ceiling of the ESPE strength.
There is an observational gap between ESPEs and SEP events directly detected during the instrumental era \cite{usoskin_gap_21}.
The gap is caused by the high detection threshold of the cosmogenic-isotope method, on one hand, and the absence of extreme events during the past decades, on the other hand. 
There might be intermediate events, but they are not detectable with the present level of measurement accuracy.
Weaker events are likely to be found later, with the progress in the isotopic measurements using acceleration mass spectrometry (AMS) \cite{guettler13}, filling the observational gap.

Such strong events remained unnoticed by ancient societies, since Earth's strong magnetic field and thick atmosphere protect us on the ground from the harmful radiation. 
They can, however, have dramatic impacts on the technological aspect of our modern civilisation, which heavily relies upon space-based technologies.
Should such an event occur nowadays, the spacecraft fleet could be instantly destroyed by corpuscular radiation, with probably an exception for those at low-altitude low-inclination orbit \cite{miyake19,usoskin_SSR_23}. 
The impact can be much stronger during weakening of the geomagnetic field, known as geomagnetic excursions or reversals, when the geomagnetic shielding is greatly reduced \cite{gao22}. 
They occur seldom and irregularly, several per million years, but last sufficiently long, a few millennia, to accommodate several ESPEs.

%=================%=================%=================%=================%=================
\section{Superflares on Sun-like stars}
\label{S:Stellar}

The cosmogenic-isotope records thus provide the only empirical evidence for extreme solar particle events, and only a handful of them are known. 
Since the period of direct solar observations spans only a few decades, it is also far too short to establish reliable statistics for the most energetic events.
To place these rare solar events in a broader context, we now consider the statistics of stellar superflares, which use ensemble monitoring of Sun-like stars to extend the effective timescale well beyond a single-star record. Stellar observations provide a valuable complementary approach: by monitoring tens of thousands of stars simultaneously, one can obtain a statistical sampling equivalent to observing a single star~-- such as the Sun~-- over hundreds of thousands of years.
Thanks to space missions such as \textit{Kepler} and \textit{Transiting Exoplanet Survey Satellite (TESS)}, which have measured brightness variations for hundreds of thousands of main-sequence stars, this approach has become feasible.  
Quantifying how often superflares occur on stars comparable to the Sun is essential for assessing the likelihood and potential impact of extreme solar activity.

\subsection{Discovery and interpretation}
\label{sec:discover}
Superflares are highly energetic analogues of solar flares (see Sect.~\ref{S:Solar}\ref{sec:flares}), releasing several orders of magnitude more energy than the largest directly observed solar flares.  Early ground-based observations detected candidates on nine F8--G8 main-sequence stars \cite{Schaefer2000}.
Subsequent broadband photometric observations of the \textit{Kepler} telescope confirmed their reality by revealing impulsive, flare-like brightenings (``white-light'' flares) on G-type dwarfs \cite{Maehara2012, Shibayama2013}.   Their peak amplitudes can reach a few percent of the total stellar radiative output, and total energies span from $\approx 10^{33}$~erg up to $\sim10^{36}$~erg \cite{Tu2020, Maehara2015, Namekata2017, Notsu2019, Okamoto2021, Vasilyev2024}.  Earlier hypotheses proposed that such energetic outbursts might be triggered by magnetic interactions with close-in planets or binary companions \cite{Rubenstein2000, Ilin2025}, but high-precision photometry and spectroscopic follow-up observations have ruled out these external scenarios as the dominant cause \cite{Notsu2019,Vasilyev2024}. Instead, observations show that superflares occur preferentially on rapidly rotating stars with large starspot coverage and enhanced chromospheric activity, indicating that strong magnetic fields are the primary driver \cite{Shibayama2013,Notsu2019,Karoff2016}. 

Here we focus on \emph{Sun-like stars}.
We note that to describe stars similar to the Sun, the terms ``Sun-like'' and ``solar-type'' are used interchangeably in the literature, sometimes referring to samples of different breadth or selection criteria (see also Sect.~\ref{S:Stellar}\ref{sec:OcRa}).
Without entering this terminological discussion, we use the term ``Sun-like stars'' to denote
G-type main-sequence dwarfs with parameters close to the present Sun: effective temperatures within $T_\mathrm{eff}=5500-6000$\,K, rotation periods in the range $P_\mathrm{rot}=20$–$30$\,d, and ages of roughly $3$–$6$\,Gyr. 
This definition is used consistently throughout the rest of this paper.

Superflares on Sun-like stars show statistical properties remarkably similar to solar flares.
Their occurrence distribution follows a power law, implying scale-free energy release \cite{Maehara2012, Aschwanden2021, Vasilyev2024} and their flare durations increase with energy with a similar slope across solar and stellar regimes, consistent with reconnection and loop cooling times scaled with event size \cite{Maehara2015,Namekata2017}. White-light emission on both the Sun and stars contains a hot, blue continuum often approximated by a $\sim$9000–10000\,K blackbody plus Balmer continuum, with strong Balmer and Ca\,II line emission \cite{Kretzschmar2011, Kowalski2024}. These similarities support a unified picture in which solar flares and stellar superflares differ primarily in scale.

\subsection{Estimates of the occurrence rates}
\label{sec:OcRa}

Table~\ref{tab:super_table} summarises key studies that estimated occurrence rates of superflares on Sun-like stars. The results are rather inconsistent. Inferring the superflare occurrence rate (sometimes called flare frequency in the literature) distribution and projecting it to the Sun requires addressing two main challenges: (i) developing robust flare-detection methods and (ii) constructing stellar samples that are representative analogues of the Sun. 

\begin{table*}[t]
\centering
\caption{Summary of key studies of superflares on Sun-like stars. The \emph{Stellar sample} column lists the selection criteria: effective temperature $T_\mathrm{eff}$, rotation period $P_\mathrm{rot}$, and the stellar variability metric $R_{\mathrm{var}}$, which is defined as the difference between the 95th and 5th percentiles of the sorted differential intensities of the light curve, computed separately for each \textit{Kepler} observing quarter \cite{Basri2010,Basri2011,Reinhold2020}. The third column reports the number of analysed stars and the observational time span for individual stars ($T_\mathrm{obs}$). Flare occurrence rates are given for events with energies $E \gtrsim 10^{34}\,\mathrm{erg}$. All studies use data from the \textit{Kepler} mission. 
}
\label{tab:super_table}
\begin{tabularx}{\textwidth}{@{} l C C r C @{}}
\toprule
\shortstack[m]{\textbf{Study}} & \textbf{Stellar sample} &  \shortstack[m]{\textbf{\# of stars} \\ \textbf{monitored}}    & \shortstack[m]{\textbf{\# of} \\ \textbf{flares}}
 & \shortstack[m]{\textbf{Flare} \\ \textbf{occurrence rate}} \\
\midrule
Maehara et al.\ (2012) \cite{Maehara2012} &  \mbox{ {$T_\mathrm{eff}=5600-6000$K}} {$P_\mathrm{rot} \ge 10$\,d} & \mbox{  {14000 }}, {$T_\mathrm{obs}=120$\,d } & 14 & $\sim$1/800\,yr$^{-1}$ \\
\bottomrule
Shibayama et al.\ (2013) \cite{Shibayama2013} & \mbox{ {$5600-6000$\,K }} {$\ge 10$\,d } & \mbox{{14325 }}, {$500$\,d } & 44 & $\sim$1/800\,yr$^{-1}$ \\
\bottomrule
Maehara et al.\ (2015) \cite{Maehara2015} & \mbox{ {$5600-6000$\,K }} {$10-20$\,d}& \mbox{{1547 }}, {$4$\,yr  }& 187 & $\sim$1/500\,yr$^{-1}$ \\
\bottomrule
Okamoto et al.\ (2021) \cite{Okamoto2021} & \mbox{ {$5600-6000$\,K }} {$20-40$\,d}& \mbox{{1641 }}, {$4$\,yr }& 26 & $\sim$1/6000\,yr$^{-1}$ ($\sim$1/3000\,yr$^{-1}$ for $7 \times 10^{33}$ erg)\\
\bottomrule
Vasilyev et al.\ (2024)  \cite{Vasilyev2024} & \mbox{ {$5500-6000$\,K }} {$20-30$\,d} $R_{\mathrm{var}}<0.3\%$ & \mbox{{5959 }}, {$4$\,yr } & 238 & $\sim$1/130\,yr$^{-1}$ \\
\bottomrule
\end{tabularx}
\end{table*}

Standard flare detection relies on identifying the characteristic fast-rise/slow-decay profile in broadband stellar light curves -- a rapid impulsive brightening followed by an approximately exponential decay \cite{Davenport2014}. This approach was widely used to analyse \textit{Kepler} data \cite{Maehara2012, Shibayama2013, Okamoto2021}. However, because light curves are extracted by summing flux within a fixed photometric aperture, a flare from a faint neighbouring star can readily be misattributed to the target star.  This risk is amplified by \textit{Kepler’s} relatively coarse pixel scale of 3.98~arcseconds, which often encompasses multiple stars.  This can introduce biases in the superflare statistics. To reduce contamination, some studies restricted their analyses to isolated targets, but such conservative criteria dramatically reduce sample sizes, excluding the majority of stars. For example, the authors of \cite{Shibayama2013} removed $\approx70$\% of the initial sample. 
For Sun-like stars, which are expected to produce superflares rarely,  retaining every reliable detection is essential for robust statistics. This is achieved by combining photometric flare identification with image-level diagnostics: candidates are flagged in the light curve and then localised on images via point-spread-function (PSF) fitting at sub-pixel spatial resolution \cite{Vasilyev2022}. Applied to \textit{Kepler} data, this technique robustly discriminates on-target events from nearby contaminants, even when several field stars fall within the same photometric aperture. 

With reliable flare detection at hand, the next step is to construct a stellar sample representative of the Sun. 
A standard approach begins with mission target catalogues (e.g., the Kepler Input Catalog \cite{Brown2011} or PLATO Input Catalog \cite{Montalto2021}) that provide fundamental stellar parameters, such as effective temperature, surface gravity, and metallicity. Narrow cuts around solar values in $T_\mathrm{eff}$, $\log g$, and [Fe/H] select Sun-like main-sequence G-dwarfs. Rotation periods are then measured from light curves using standard time-series diagnostics (power spectra or autocorrelation functions) \cite{McQuillan2013}, or taken from existing rotation catalogues \cite{Nielsen2013, McQuillan2014, Reinhold2023}.
Slow rotators (e.g., $P_\mathrm{rot}\gtrsim20\,\mathrm{d}$) are retained to match the contemporary Sun ($\sim$25\,d) and to avoid the high-activity regime typical of young, fast rotators.  
Fast rotators are more active and flare much more frequently,  which would bias the inferred superflare rate for the Sun. Finally, astrometric information (e.g., from \textit{Gaia}) and binary-star catalogues are used to exclude close binaries and obvious blends, yielding Sun-like, slowly rotating, apparently single stars \cite{Vasilyev2024}. 

However, determining reliable rotation periods for slowly rotating stars such as the Sun presents an additional challenge. On such stars, most starspots evolve or disappear within timescales comparable to the stellar rotation period \cite{DrielGesztelyi2015}, so their brightness variations do not repeat regularly from one rotation to the next.
This leads to light curves that appear irregular or “aperiodic”, where the rotational signal is weak or even absent.
In addition, brightness variations due to dark spots are partly compensated by bright faculae (see Sect.~\ref{S:Solar}\ref{sec:ARs}), which reduces the amplitude of solar brightness variation 
and lowers the signal-to-noise of rotational modulation.  
This facular compensation becomes particularly important for slowly rotating, weakly active stars like the Sun, where most of the magnetic flux emerges in small-scale elements rather than large spot groups \cite{Thornton2011,Parnell2009,Hofer2024}.
With increasing magnetic activity, a larger fraction of the flux emerges in big, dark spots while the facular-to-spot area ratio decreases \cite{Foukal1994,Shapiro2014,Yeo2021,Krivova2021}, so brightness modulation is weaker for less active, slower-rotating stars.
Since the resulting photometric variability on Sun-like stars is weak, rotation-period recovery becomes difficult and also highly sensitive to secondary factors such as inclination, spot evolution, facular contributions, and metallicity \cite{Shapiro2017, Eliana2020,Witzke2020,Reinhold2021}. For example, when the Sun is treated as a \textit{Kepler} target, the standard autocorrelation-function method recovers the correct rotation period in only  $\sim$3\% of cases across varying viewing inclinations, metallicities, and magnitude-dependent noise levels \cite{Reinhold2021}. 
Consequently, by selecting only stars with known rotational periods~-- the latter requiring a clear, periodic rotational signal~-- past stellar samples were biased towards stars with large, stable spots.
Such stars are, however, typically more magnetically active and photometrically variable than the Sun \cite{Reinhold2020}.
Consistent with this, an analysis of super-flaring stars in the \cite{Okamoto2021} sample found that only one of the 265 Sun-like stars was directly comparable to the present-day Sun \cite{Herbst2025}.
This bias towards higher activity explains why broad stellar samples, which include many faster rotators (e.g., \cite{Maehara2012,Shibayama2013,Maehara2015}; $10 \lesssim P_\mathrm{rot} \lesssim 20\,\mathrm{d}$), 
tend to overestimate superflare occurrence rates and cannot be applied directly to the current Sun (see also Materials and Methods in \cite{Vasilyev2024}).

\begin{figure}
    \centering
    \includegraphics[width=1.0\linewidth]{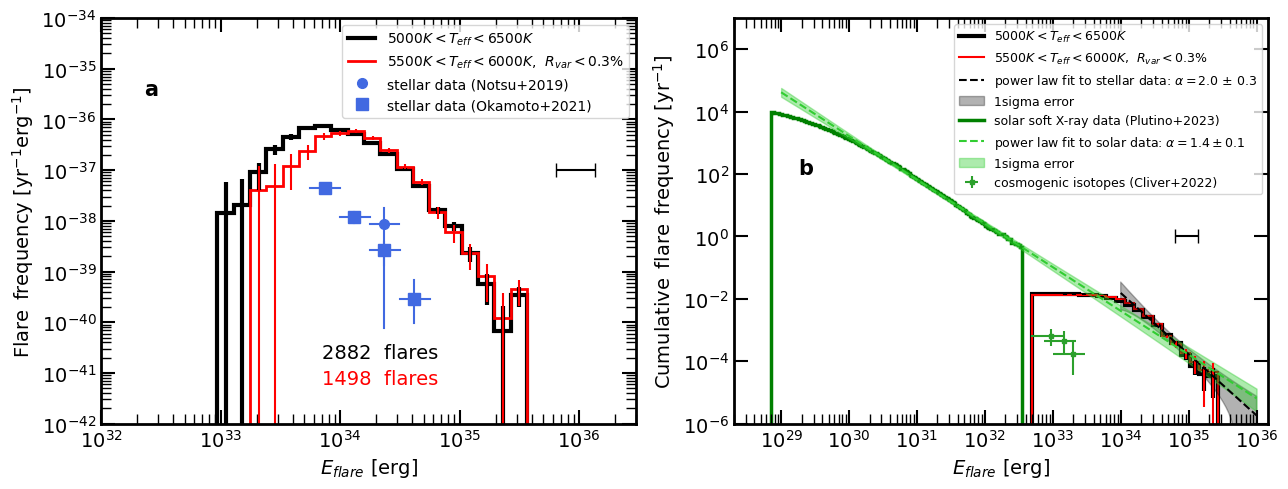}
\caption{\textbf{Occurrence rate vs. energy distributions of solar and stellar flares.} 
\textbf{(a)} Flare occurrence rate per star per year per unit energy for all stars in the \textit{Kepler} sample (black) and for a restricted subset of Sun-like stars with effective temperatures of 5500--6000\,K and photometric variability $R_\mathrm{var} < 0.3\%$ (red). Blue symbols show representative results from earlier studies \cite{Notsu2019,Okamoto2021}. 
\textbf{(b)} Cumulative flare occurrence rate (number of flares per star per year above a given energy $E$) for the two stellar samples shown in panel~\textbf{(a)} (red and black curves). The black dashed line represents a power-law fit to the distribution of stellar flares with $E \gtrsim 10^{34}$\,erg. For comparison, the green curve shows the solar flare occurrence rate distribution derived from soft X-ray observations of 334,122 flares between 1986 and 2020 \cite{Plutino2023}. Green symbols mark the inferred frequencies of ESPEs inferred from cosmogenic isotope records  \cite{Cliver2022}. The horizontal error bar below the legend indicates the mean uncertainty in stellar flare energy. Figure adapted from \cite{Vasilyev2024}.}
    \label{fig:valer_solar_stellar_frequnecy_distribution}
\end{figure}

To overcome these biases, \cite{Vasilyev2024} constructed a Sun-like sample (see Sect.~\ref{S:Stellar}\ref{sec:discover}) matched not only in $T_\mathrm{eff}$ and $P_\mathrm{rot}$ but also in photometric variability comparable to the Sun.
They derived an occurrence rate of roughly one event per century for $E \gtrsim 10^{34}$\,erg (see Figure~\ref{fig:valer_solar_stellar_frequnecy_distribution}).  
Together, these findings provide the most consistent framework to date for linking solar flare and stellar flare statistics.  

\subsection{Summary}
High-precision space-based photometric surveys show that Sun-like stars can produce white-light flares with energies up to $10^{34}$–$10^{35}$\,erg, albeit at low rates (century–millennium scales). When stellar samples are carefully matched to the Sun in $T_\mathrm{eff}$, rotation, and photometric variability,  and flares are confirmed as on-target, the inferred rate for events with $E\gtrsim10^{34}$\,erg is of order one per century \cite{Vasilyev2024}. 
Extrapolations of the solar flare occurrence rate vs. energy distribution align with these stellar results, suggesting that solar and stellar flares are governed by the same underlying physical mechanism. For the Sun, this implies that superflares are possible, and their absence during the space era is consistent with low occurrence rates.

From the solar perspective, an independent estimate comes from direct solar observations.
Based on GOES soft X-ray (SXR) observations, \cite{Hudson2024} derived an occurrence-frequency distribution for peak SXR flux 
using $\sim 600$ events spanning the upper $\sim$1.5~orders of magnitude of the observed SXR flare intensity distribution.
Extrapolating this distribution suggests a lower limit return time of 
$\gtrsim10^{6}$\,yr for flares with bolometric energies $\sim 10^{34}$~erg. 
However, the two approaches rely on different observables (SXR peak flux versus bolometric energy), vastly different statistical baselines ($\sim 50$~yr for one star versus a few years for tens of thousands of stars), and distinct degrees of extrapolation beyond their direct empirical coverage. 
Consequently, the two estimates are not directly comparable.

Future progress requires both methodological and observational advances. High-resolution spectroscopy of superflare stars can constrain stellar ages,  activity levels,  and help identify and reject unresolved binaries. On the observational side, upcoming space missions such as \textit{PLATO} \cite{PLATO} will provide high-precision, long-duration light curves for bright Sun-like stars, dramatically expanding the accessible sample and enabling detailed synergy with spectroscopy. Together, these advances will refine occurrence-rate estimates, clarify the physical mechanisms of superflares, and strengthen the link between solar and stellar extreme activity.

The comparison between solar and stellar flares suggests that both are governed by the same fundamental process operating across many orders of magnitude in energy. 
To understand what ultimately limits the energy of solar eruptions and how these limits relate to the occurrence of superflares on Sun-like stars, we next discuss the emergence and evolution of the solar magnetic field, eventually leading to eruptive activity.

%=========================
\section{Solar magnetism and energetic events}
\label{S:Solar}
Extreme Solar Particle Events are among the most powerful manifestations of solar activity.
They are typically associated with large coronal mass ejections (CMEs), which are vast eruptions of magnetised plasma expelled from the solar corona into interplanetary space. CMEs are a key driver of space weather throughout the heliosphere.

However, CMEs themselves do not arise in isolation.
They are often, though not always, accompanied by solar flares.
Both phenomena result from the rapid reconfiguration of magnetic fields in the solar atmosphere through reconnection, with flares releasing energy mainly as radiation and energetic particles, and CMEs carrying it away as kinetic and magnetic energy \cite{Shibata2011}.
Although CME-driven shocks are generally considered the dominant accelerators of the high-energy particles observed during large SEP events, flares can also accelerate particles directly in the reconnection region or along magnetic loops  \cite{desai_LR_16,Reames2013}. The ground-level enhancement of January 20 2005 (GLE\#69; see Sect.~\ref{S:ESPE}\ref{sec:ESPE-discover}) is a well-documented case where relativistic particles appear to have been accelerated primarily by the flare itself~-- an example of a so-called impulsive SEP event \cite{Grechnev2008,Klein2014}.

Flares, in turn, are rooted in active regions (ARs)~-- concentrations of magnetic flux on the solar surface that manifest themselves as sunspots, faculae or plage (plage is a chromospheric counterpart of faculae observed in the photosphere).
These regions are visible traces of the solar magnetic field generated by the internal dynamo.
As magnetic flux emerges and evolves, it can store large amounts of free magnetic energy in stressed or sheared coronal fields, which is eventually released through reconnection, driving flares, CMEs, and SEP events \cite{Priest2002,Toriumi2019}.

The surface magnetic field is therefore the fundamental driver of all energetic solar phenomena.
Understanding how magnetic flux emerges, accumulates, and reconnects is essential for linking solar magnetism to superflares and extreme solar particle events.

\subsection{Magnetic field as the origin of solar activity}
\label{sec:mf}

The magnetic field of the Sun underlies all manifestations of solar activity.
It is generated by a dynamo operating in the convection zone, where differential rotation and convective motions convert kinetic energy into magnetic energy \cite[e.g.,][]{Charbonneau2010}.
The resulting large-scale toroidal fields become buoyant, rise through the convection zone, and emerge at the surface as bipolar active regions, visible as sunspots, faculae and plage.

The emergence, evolution, and decay of these magnetic structures determine the amount of free magnetic energy available for eruptions.
The gradual build-up and reconfiguration of these magnetic fields in the photosphere and corona can later release energy through reconnection, giving rise to flares and coronal mass ejections \cite[e.g.,][]{Priest2002,Toriumi2019}.
Consequently, the rate and magnitude of eruptive events are ultimately determined by the amount of flux emerging from the solar interior.

Variations in solar magnetism on various timescales shape the heliospheric environment and, in particular, modulate the flux of galactic cosmic rays reaching Earth.
This modulation, in turn, affects the production of cosmogenic isotopes such as $^{14}$C, $^{10}$Be and $^{16}$Cl (Table~\ref{tab:IU-isotopes}) in the atmosphere, which provide an indirect record of long-term solar magnetic activity \cite{wu18,usoskin_LR_23}.
Thus, the magnetic field generated by the solar dynamo links the interior processes of the Sun to heliospheric variability and solar-terrestrial effects.

\subsection{Active regions: magnetic complexity and energy storage}
\label{sec:ARs}

The magnetic field generated by the solar dynamo emerges at the surface in bundles that form active regions. 
These regions contain concentrations of opposite magnetic polarity and typically contain sunspots, faculae and plages. 
During the emergence process through the convective zone and interactions with surface flows \cite[see][]{Cheung2014}, the magnetic field is sheared and twisted, leading to non-potential configurations storing large amounts of free magnetic energy in the corona.

Larger active regions can store more magnetic energy.
However, size alone does not determine the eruptive potential of an AR.
Another key factor is the magnetic complexity of the regions \cite{Toriumi2019}.
Statistical studies have shown that flare productivity increases with the total unsigned magnetic flux, area, and complexity of an AR \cite{Schrijver2007,Tiwari2010}. 

It is also notable that magnetic flux emergence tends to cluster in time and space~-- a phenomenon commonly referred to as ‘‘nesting’’.
Nesting implies that new flux is more likely to emerge near existing active regions than at random locations on the disc \cite{Gaizauskas1983,Isik2020_nesting}.
Some authors interpret long-lived concentrations of activity as ‘‘active longitudes’’\citep[e.g.][]{berdyugina03}, although the existence and physical nature of persistent active longitudes on the Sun and stars remain debated.
Such clustering can enhance both the local magnetic flux budget and the complexity of an active region, thereby increasing the probability of extreme events.
A recent example is NOAA AR~13664, which emerged in late April 2024 and produced multiple X- and M-class flares and coronal mass ejections in May.
This active region had a complex structure and evolution, with one or maybe even two further ARs emerging and merging into it \cite{Dikpati25,kontogiannis2025}.

\subsection{Solar flares}
\label{sec:flares}
Magnetic reconnection in active regions reconfigures the field topology and triggers a rapid release of free magnetic energy \cite[see][for a review]{Shibata2011}.
Two complementary manifestations of this process are solar flares and coronal mass ejections: the former radiate the released energy and accelerate particles in the lower corona, whereas the latter expel mass and magnetic flux into interplanetary space.
Although closely related, they do not always occur together.

The total flare energy is governed by the amount of magnetic flux participating in reconnection.
Typical large solar flares release bolometric energies in the range 
$10^{30}-10^{32}$~erg, while estimates for the most extreme historical event~--- the Carrington flare of 1859~--- approach a few times $10^{33}$~erg \cite{hayakawa_magnitude_2023,Hudson2025}.
Direct measurements of total solar irradiance during the powerful (X17 and X28+) flares of the Halloween event in October--November 2003 indicate that these flares reached bolometric energies of about $(2-6)\times 10^{32}$~erg \cite{woods_contributions_2006, Emslie2012, Moore2014}. 
Three-dimensional Magnetohydrodynamic (MHD) simulations scaled to the largest sunspot group observed in 1947 yielded flare energies up to about $6\times 10^{33}$~erg \citep{Aulanier2013} as a conservative theoretical upper limit.
Independent estimates based on magnetic flux budgets of active regions and historical records place the upper limit below $\sim10^{34}$~erg \cite[see][for a review]{Cliver2022}.

However, empirical scaling based on observed active-region and flare ribbon properties suggests that the most extreme directly documented active regions had, in principle, the potential to produce flares of at least several times $10^{34}$~erg \cite{Krivova2025}.
Flare ribbons are elongated bright structures in the chromosphere that mark the footpoints of newly reconnected coronal field lines.
They provide a useful proxy for the amount of reconnected magnetic flux~--- and thus for the released energy.
Using observations from the Solar Dynamics Observatory \cite[SDO;][]{pesnell_solar_2012} of more than 300 large flares between 2010 and 2016, the authors of \cite{Kazachenko2017} established empirical relationships between ribbon area, total reconnection flux, peak X-ray flux, and thus eventually the total flare energy.
Their catalogue also lists the total area of the associated active regions, which can in turn be estimated from the observed areas of sunspots.
Relationships derived from the catalogue \cite{Kazachenko2017} can be extrapolated to estimate flare energies for the largest historical active regions \cite{Krivova2025}.
Estimates based on these relationships of the expected energies for the Halloween (2003) and Bastille Day (July 2000) events agree with observations. 
Applying this scaling to the largest historical active regions considered individually (i.e. without accounting for nesting) yields possible flare energies of several $10^{34}$~erg for the great sunspot of 8 April 1947, whose area reached over 6100~msh (millionths of the solar hemisphere) \cite{mandal_sunspot_2020}.
AR nesting (see Sect.~\ref{S:Solar}\ref{sec:ARs}) could increase this potential even further.
For comparison, the measured areas of sunspots forming the cores of ARs responsible for the Carrington (1859) and Halloween (2003) events were ``only'' about 3100--3500~msh \cite{hayakawa_magnitude_2023,meadows_size_2024,ermolli_solar_2023} and 3500~msh \cite{mandal_sunspot_2020}, respectively.
Assuming the same empirical scaling and spot area 3100~msh, \cite{Krivova2025} estimated that the bolometric energy of the Carrington flare could have reached about $7\times10^{33}$~erg, consistent with independent analyses of historical records \cite{hayakawa_magnitude_2023,Hudson2025}.
These estimates suggest that under favourable conditions~-- particularly in large or nested active regions~-- solar flares could reach energies of several $10^{34}$~erg, approaching the lower end of the stellar superflare regime.

Whereas solar flares are most easily detected in X-rays, energetically they are dominated by their UV–optical continuum output.
However, due to the lack of appropriate observations, especially in the ultraviolet, the spectral energy distribution of flare emission remains poorly constrained. 
Observational and modelling studies indicate that the radiative energy emitted in hard X-rays and $\gamma$-rays ($<$0.1~nm) constitutes less than 1\% of the total, while soft X-rays (0.1--10~nm) add perhaps another roughly 1\% \cite[e.g.][]{woods_contributions_2006,emslie_global_2012}.
Most flare radiation is emitted at longer wavelengths: roughly 40–60\% below 190~nm (extreme and far-ultraviolet) and 40–60\% above 190 nm (near-UV, visible and near-infrared) \citep{woods_contributions_2006,kretzschmar_sun_2011}.
The gap in our knowledge of the flare budget partition emphasises the importance of broad-band observations for constraining the total flare energy budget, for example, such as those expected from the Solar Ultraviolet Imaging Telescope (SUIT) onboard Aditya-L1 mission \cite{Ghosh_SUIT-2016,Tripathi_SUIT-2025}.

\subsection{Coronal mass ejections}
\label{sec:CMEs}

Coronal mass ejections are large-scale eruptions of magnetised plasma from the solar corona into interplanetary space.
They can carry up to $10^{16}$~g of material and kinetic energies exceeding $10^{32}$~erg, and are a major driver of heliospheric disturbances and space weather \cite{Vourlidas2010,Webb2012,Kilpua_LRSP}. 
Flares and CMEs are both manifestations of magnetic reconnection and the release of stored magnetic free energy, but the relationship between them is complex.
Not all large flares produce CMEs, and not all CMEs are accompanied by major flares \citep[e.g.][]{Andrews2003,Yashiro2005}. 
Flares accompanied by CMEs are usually called eruptive, while those without are termed confined.
The likelihood of CME association increases with flare magnitude: above roughly X4--X5, nearly all flares are eruptive \cite{Yashiro2005}.

CMEs and flares share the same magnetic energy reservoir.  
In eruptive events, a substantial fraction (often more than half) of the released energy is channelled into CME kinetic and potential energy, while the remainder is partitioned between flare radiation, heating, and accelerated particles \citep{Yashiro2005,emslie_global_2012}.  
This energy partition also governs particle acceleration: it is most efficient when reconnection occurs in events that open magnetic field lines~-- typically those accompanied by CMEs~-- whereas confined flares, despite their strong radiation, have limited escape channels for energetic particles.
In confined flares, where overlying magnetic fields prevent large-scale ejection, most of the released energy remains trapped in the lower atmosphere \citep{Thalmann2015,Sun2015,Toriumi2017,Li2020}. 

Nevertheless, although CME-driven shocks dominate the production of high-intensity gradual SEP events, flare reconnection can also accelerate relativistic particles.
A notable example is GLE~\#69 of 20~January 2005, where the earliest, hard-spectrum particle component was flare-associated, even though the event was accompanied by a very fast CME \citep{Grechnev2008,Klein2014,Gopalswamy2018}.
This demonstrates that the early flare-accelerated SEP component can contribute significantly under suitable coronal and interplanetary conditions, complementing the later shock-accelerated component.

Overall, eruptive flares with fast CMEs tend to produce the most intense SEP events, whereas confined flares may appear relatively bright but usually yield fewer escaping particles.

This effect might have implications for comparisons to stellar superflares.  
Numerical simulations suggest that on active stars, the overlying coronal fields can inhibit CME escape for flare energies exceeding $10^{34}$–$10^{35}$~erg \citep{Drake2013,AlvaradoGomez2018}.
As a result, superflares may occur without accompanying CMEs, releasing their energy predominantly as radiation rather than mass ejection \cite{Namekata22}.  
This could explain why many stellar superflares show no evidence for large-scale mass loss, highlighting the crucial role of magnetic confinement in shaping the energy partition of solar and stellar eruptions, see also \citep{Toriumi2017,Odert2017}.

\subsection{Solar energetic particles}

Solar energetic particles are formed by a non-thermal population of charged particles ultimately originating from the Sun.
The bulk of SEPs has energy in the non-relativistic range of a few tens of MeV per nucleon, but sometimes they can be accelerated up to relativistic energy of several GeV/nucleon.
Formation of SEPs requires an effective acceleration mechanism, which can operate in the solar conditions only rarely, during eruptive events such as flares or CMEs.
Accordingly, SEPs are not omnipresent in the Earth's neighbourhood, but rather form sporadic SEP events.

Two main mechanisms are known to accelerate charged particles to make them SEPs \cite{cane86,vainio09}.
One is the direct acceleration by the electric field during magnetic reconnection in the solar atmosphere during solar flares.
This is a very effective and fast mechanism, but it has difficulties in producing strong fluxes of SEPs -- the time and volume of the reconnection are limited.
Moreover, the conditions should be favourable, viz, open magnetic field configuration, to let these accelerated particles escape from the Sun and become SEPs \cite{masson19}.
This type of flare-related acceleration typically produces `impulsive', that is, short, intense and hard-spectrum SEP events.
The typical duration of impulsive events is tens of minutes -- hours.
Impulsive SEP events are often anisotropic, with the axis of anisotropy being aligned with the direction of the interplanetary magnetic field line. 

The other mechanism is related to shocks driven by fast CMEs in the corona and interplanetary medium \cite{desai_LR_16}.
The shock acceleration is not as fast as reconnection, but takes a longer time and involves large volumes, and thus, can produce strong SEP events.
Because of their longer duration, several hours to a day, such events are called `gradual' and often have a softer spectrum.
This mechanism requires a seed population of non-thermal pre-accelerated particles.
Gradual SEP events are typically more isotropic.
For strong SEP events, both components, viz., the impulsive (prompt) and gradual (delayed), are often observed within the same event, implying that both mechanisms may work subsequently.

SEPs are charged particles and are thus subjected to a complex transport in the interplanetary magnetic field before reaching Earth.
This transport includes escape from the acceleration region in the corona, diffusive propagation along the magnetic field lines in the interplanetary space, and focusing in the diverging magnetic field.
When the flux of energetic particles is high, they can generate plasma waves near the acceleration sites and scatter away, reducing the acceleration efficiency.
This effect is known as `streaming limit' and has been observed for low-energy ($<$100 MeV) SEPs \cite{lario08,reames17}, but it may potentially affect even higher energies for extreme events. 
The heliospheric current sheet can also play a role in the interplanetary transport of SEPs \cite{waterfall22}.
At present, a full self-consistent modelling of such ESPEs from acceleration at/near the Sun to their impinging on Earth is impossible, but the work in this direction is in progress.

These complex processes of energetic particle acceleration and transport make the relation between SEP events and flares unclear.
While the relation is statistically significant for a large number of weak and moderate events \cite{papaioannou23}, its projection to strong and extreme events remains elusive \cite{Cliver2022}. 
This can be readily seen in the fact that the strongest SEP events were not caused by the strongest flares and vice versa.
This reflects the dual dependence of particle production on both flare reconnection energy and CME-driven shock efficiency.  
The most extreme solar particle events likely require both components~--- strong flares to provide seed particles and magnetic connectivity, and powerful CMEs to accelerate them efficiently and allow their escape into interplanetary space.
This is further addressed in Section~\ref{S:relation}.

\subsection{Summary}

Solar surface magnetism~-- its emergence, evolution, and reconnection~-- provides the physical foundation linking active regions, flares, CMEs, and ultimately SEPs and ESPEs.
Solar energetic phenomena are driven by the magnetic field generated by the solar dynamo and its complex evolution through the convection zone and atmosphere. 
Active regions act as magnetic reservoirs where free energy accumulates through shear, twist, and flux emergence. 
When the stored energy is released by magnetic reconnection, it produces flares, coronal mass ejections, and energetic particle acceleration.

Flares and CMEs represent complementary channels of magnetic energy conversion: radiation and particle acceleration in the former, and bulk kinetic energy and magnetic flux transport in the latter. 
Their relative partition depends on the surrounding coronal magnetic topology. 
In confined events, strong overlying fields trap the released energy in the lower atmosphere, enhancing radiative output, whereas eruptive flares divert a substantial fraction into CME kinetic energy. 
This distinction may help to explain why some of the most powerful stellar superflares appear to lack clear CME signatures: strong overlying fields could prevent mass ejection while allowing intense radiative flaring.

Energetic particles accelerated in solar flares and CME-driven shocks form the population of solar energetic particles, whose most extreme manifestations are the ESPEs observed in cosmogenic isotope records. 
Thus, the entire chain~-- from dynamo action and magnetic flux emergence to reconnection and particle acceleration~-- connects the magnetic life cycle of the Sun to its most energetic transients. 
Understanding this coupling provides the physical basis for extrapolating solar behaviour towards the superflare regime and assessing the plausible upper limits of solar and stellar particle events.

%==============================
\section{Discussion and Summary: Are ESPEs and superflares related?}
\label{S:relation}
Extreme Solar Particle Events and stellar superflares represent the most energetic manifestations of magnetic activity on the Sun and Sun-like stars.
ESPEs involve particle fluences comparable to those accumulated over a millennium of ordinary solar activity, whereas superflares release radiative energies far exceeding any observed on the Sun.
Both are extremely energetic eruptive events, occurring only rarely, yet it remains uncertain whether they are two sides of the same coin~-- that is, directly related phenomena.
The fact that these two types of events have been discovered and are studied by different methods on different objects complicates this topic.
ESPEs are known from indirect proxies for one star, our Sun, on the timescale of about 15~millennia, while superflares are studied by using space-borne telescope observations of thousands of distant stars during only a 4-year period.
Although the statistics of the two methods are comparable, a few tens of thousands of star-years, a direct comparison of the single-object time variability to a large ensemble average is not yet fully validated.

Three hypotheses can be made to compare the statistics of ESPE vs. superflares (SF):
\begin{enumerate}
\item Supeflares and ESPEs are nearly one-to-one related (ESPE $\Leftrightarrow$ SF), so that their occurrence rates can be directly compared. 
\item ESPEs are produced by superflares but only rarely, so that an ESPE is caused by a superflare, but not every superflare leads to an ESPE. (SF $\Rightarrow$ ESPE). 
\item ESPEs and flares are not directly related (SF $\nLeftrightarrow$ ESPE), thus, their statistics are incomparable. 
\end{enumerate}
The first hypothesis was assumed as the null hypothesis in earlier works, when the uncertainties in the statistical results for both types of events were big enough to accommodate it. 
However, recent analyses demonstrated that superflares are significantly more frequent on Sun-like stars than ESPEs on the Sun \cite{Vasilyev2024}.
Moreover, all efforts to find a signature of an SPE signal in cosmogenic isotopes for the strongest known solar flare, the Carrington flare on 1st September 1859, failed -- there was no detectable ESPE related to the Carrington flare \cite{Miyake2023,uusitalo24}.
This rejected hypothesis (i) as very unlikely, making hypothesis (ii) the current paradigm in that ESPEs are produced by superflares, but not every superflare leads to an ESPE (e.g., \cite{Cliver2022,Vasilyev2024}), only when favourable conditions occur.
This complicates direct comparison by involving an unknown probability of an ESPE being produced by a superflare.
However, the known statistics of directly observed strong solar flares and SPEs suggest that the relation between them is poor -- the strongest SPEs were not associated with the strongest flares and vice versa \cite{koldobskiy25}. 
Extrapolating, in a probabilistic way, this statistical relation to extreme energies, these authors concluded that ESPEs are more likely to be produced by strong but not yet extreme flares in the X1–X10 range, with superflares generating ESPEs only rarely when SEP acceleration conditions are highly favourable. 
This forms hypothesis (iii), which assumes that the statistics of ESPEs and superflares cannot be directly comparable.
While hypothesis (i) is rejected, with high confidence, by the observed data, we cannot distinguish between hypotheses (ii) and (iii).
Accordingly, although the processes of the eruptive energy releases and particle acceleration during a solar flare may be physically related, the combined effects of acceleration efficiency, magnetic topology, and interplanetary transport make the direct link between flare energy and ESPE strength elusive.
Systematic studies of ESPEs in cosmogenic-isotope records on millennial timescales paired with a large ensemble of carefully selected samples of Sun-like stars can shed new light on the physics of extremely energetic processes on the Sun and cool stars.

A speculative but physically plausible possibility concerns how 
the balance between flare radiative output, CME occurrence and particle escape changes with increasing magnetic activity.
On the Sun, the probability that a flare is accompanied by a CME increases with flare energy \citep{Yashiro2005}, yet 
strong overlying fields can sometimes prevent mass ejection even in powerful events
\citep{Sun2015,Thalmann2015,Li2020}.

Solar MHD simulations suggest that such confinement may already become effective for flares with bolometric energies above a few $\times 10^{32}$~erg under present-day solar conditions \citep{AlvaradoGomez2018}.
In this regime, magnetic reconnection can still produce very energetic flares, while CME escape and efficient particle release are suppressed.
If the Sun operates near this confinement threshold, this could help explain why solar flare energies may occasionally approach $\sim10^{34}$~erg, as suggested by stellar statistics and empirical estimates of the potential energy stored in the largest historical active regions, while ESPEs remain far rarer:
as magnetic complexity increases, the likelihood of efficient particle escape may decline more rapidly with growing flare energy.

Within this speculative scenario, the observations and physical arguments are naturally consistent with hypothesis (ii), in which superflares are capable of producing ESPEs but only under a restricted set of magnetic and heliospheric conditions.
At the same time, the present evidence does not exclude hypothesis (iii), in which ESPEs and superflares are not directly related.

\iffalse
A speculative but physically plausible possibility could be that the balance between flare radiative output and CME occurrence changes with increasing magnetic activity.
On the Sun, the probability that a flare is accompanied by a CME increases with flare energy \citep{Yashiro2005}, yet some of the most powerful events remain confined when strong overlying magnetic fields prevent mass ejection \citep{Sun2015,Thalmann2015,Li2020}.
Numerical simulations and stellar observations suggest that, on more magnetically active stars, such confinement may become even more effective, suppressing CME escape above $\sim10^{34}$–$10^{35}$ erg \citep{Drake2013,Odert2017,AlvaradoGomez2018}.
In this regime, magnetic reconnection could still produce extremely energetic superflares but without accompanying mass ejection.
This would channel most of the energy into radiation rather than kinetic outflow, consistent with the high flare energies and apparent lack of CMEs on super-flaring Sun-like stars.
\textbf{Whether an analogous confinement regime is ever realized on the present Sun is unclear. At most, one might speculate that if the Sun operates near the lower edge of this regime,} it could explain why solar flare energies may reach up to $\sim10^{34}$~erg while ESPEs remain far rarer than stellar superflares: above some range of activity and complexity, the likelihood that a flare efficiently produces an Earth-reaching particle enhancement may decline because confinement prevents effective particle escape.
\fi

\ack{ We thank the reviewers for the careful and thoughtful reading of our manuscript and helpful suggestions.
VV acknowledges support from the Max Planck Society under the grant ``PLATO Science'' and from the German Aerospace Center under ``PLATO Data Center'' grant 50OO1501.
IU acknowledges partial support from the European Research Council Synergy Grant (project 101166910) and the Research Council of Finland (project 354280). We thank the Royal Astronomical Society (RAS) for supporting our participation in the Radiocarbon and Cosmic Radiation Events meeting.}

%%%%%%%%%% Insert bibliography here %%%%%%%%%%%%%%

\bibliographystyle{RS}
\bibliography{sample}
\end{document}